\documentstyle[aps,floats,epsf,twocolumn]{revtex}

\tighten
\begin{document}
\draft

\title{Critical temperature and superfluid density suppression in disordered 
high-$T_c$ cuprate superconductors}

\author{M. Franz$^1$, C. Kallin$^{2,*}$, A. J. Berlinsky$^{2,*}$ and 
M. I. Salkola$^2$} 

\address{$^1$Department of Physics and Astronomy, Johns Hopkins University,
Baltimore, MD 21218, USA\\
$^2$Department of Physics, Stanford University, Stanford, CA 94305, USA
\\ {\rm(\today)}}
\address{~
%\begin{abstract}
\parbox{14cm}{\rm
\medskip
We argue that the standard Abrikosov-Gorkov (AG) type theory of $T_c$
in disordered $d$-wave superconductors breaks down in short coherence 
length high-$T_c$ cuprates. Numerical calculations within the 
Bogoliubov-de Gennes formalism  demonstrate that the correct 
description of such systems must allow for the spatial variation of the  order parameter, which is strongly suppressed in the vicinity of impurities
but mostly unaffected elsewhere. Suppression of $T_c$ is found
to be significantly weaker than that predicted by the AG theory, in
good agreement with experiment. 
}}
%\end{abstract} 
\maketitle

%%%%%%%%%%%%%%%%%%%%%%%%%%%%%%%%%%%%%%%%%%%%%%%%%%
\narrowtext
Sensitivity of the superfluid density and the critical temperature to 
moderate amounts of substitutional disorder served as an early indicator
of the unconventional nature of the order parameter in high-$T_c$ cuprate
superconductors. 
The theory of dirty $d$-wave superconductors, pioneered in this context
by Annett, Goldenfeld and 
Renn \cite{Annett} and by Hirschfeld and Goldenfeld\cite{Hirschfeld}, 
successfully explained the observed $T^2$ dependence of the superfluid
density $\rho_s(T)$ in YBa$_2$Cu$_3$O$_{7-\delta}$ (YBCO) films and single 
crystals, attributing it
to the effect of the finite density of states at the Fermi level induced by
impurities, treated as unitary scatterers. 
Crossover to $T$-linear behavior was predicted for lower
impurity content, which was later observed in very clean YBCO single crystals
by Hardy {\em et al.}\cite{Hardy}. The theory was subsequently refined\cite{Prohammer,KPM,Maki} to provide a good quantitative description 
of the low-temperature behavior of $\rho_s(T)$. 

\begin{figure}
\epsfxsize=8.0cm
\epsffile{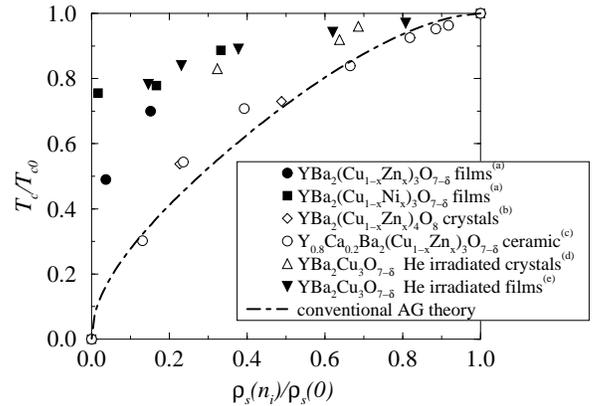}
\caption[]{Normalized critical temperature versus the normalized zero 
temperature superfluid density. Experimental data were obtained by 
$\rm ^{(a)}$mutual inductance\cite{Ulm}, $\rm ^{(b,d)}$infrared  reflectance
\cite{Basov,Basov1}, $\rm ^{(c)}$muon spin rotation\cite{Tallon}, and 
$\rm ^{(e)}$field-current density analysis\cite{Moffat}. Theoretical curve 
is from Ref.\cite{KPM}.} 
\label{fig:1}
\end{figure}

Similar models have been 
employed to predict the change of $T_c$ from its clean value $T_{c0}$ due to 
disorder\cite{KPM,Radtke}, essentially by extending the standard
Abrikosov-Gorkov (AG) theory\cite{AG} to the case of $d$-wave superconductors 
with scalar impurities. It has been noted that the experimentally observed
$T_c$ is much more robust than one would expect from these simple models,
when measured against the corresponding change in $\rho_s(0)$. Figure
\ref{fig:1} illustrates this point by showing the experimental $T_c$
versus $\rho_s(0)$ for YBCO samples disordered by different types of disorder
as obtained by various groups\cite{Ulm,Basov,Basov1,Tallon,Moffat} and 
the corresponding
theoretical prediction\cite{KPM}. While there exists a considerable spread
in the experimental data, it is quite evident that the theory systematically
overestimates the suppression of $T_c$, perhaps by as much as a factor
of 2 in the cases of Refs.\cite{Ulm,Basov1,Moffat}. In order to remedy this 
situation, more 
realistic models have been considered, taking into account strong coupling
corrections within the Eliashberg formalism\cite{Radtke} together with 
realistic band structures\cite{Arberg}, proximity of the Fermi level to a
van Hove point\cite{Fehrenbacher} and the details of the pairing interaction
within the spin-fluctuation model\cite{Pines}. While some improvement over the simple model can be achieved for carefully selected parameters, the
discrepancy between theory and experiment remains in place, suggesting
that methods traditionally employed to determine the suppression of
$T_c$ by impurities are inadequate for the high-$T_c$ cuprates.
It has also been suggested that perhaps the effects of disorder on the transition from the rather peculiar and poorly understood normal state 
cannot be satisfactorily described within the framework of a simple BCS-like
mean field theory\cite{Hirschfeld2}.

 In the present article we show that
one can, in fact, formulate an adequate description within mean-field 
theory, if the spatial variations of the order parameter caused by random
disorder are accounted for in a fully self-consistent manner. We present a 
simple argument for the breakdown of the conventional AG type theory (which
enforces a uniform gap averaged over disorder) in superconductors with 
very short coherence lengths such as the high-$T_c$ cuprates. This argument is then given substance by numerical calculations within the Bogoliubov-de 
Gennes (BdG) framework, carried out for a model $d$-wave superconductor
with a random distribution of non-magnetic point impurities. Such 
calculations convincingly demonstrate that  the rapid drop in the
predicted $T_c$ is an artifact of AG theory resulting from spatial 
averaging of the order parameter. The true critical temperature
obtained from a fully self-consistent solution of the BdG equations is 
always higher, in agreement with experiment.

In a $d$-wave superconductor non-magnetic impurities are 
pair breaking and lead to suppression of $T_c$. A simple modification 
of the original AG approach\cite{AG} yields an equation for $T_c$ of the
familiar form\cite{KPM,Radtke,Arberg}:
\begin{equation}
\ln\left({T_{c0}\over T_c}\right)=
\psi\left({1\over 2}+{\alpha T_{c0}\over 2\pi T_c}\right)-
\psi\left({1\over 2}\right),
\label{tc}
\end{equation}
where $\psi$ is a digamma function and $\alpha=1/2\tau T_{c0}$ is a pair
breaking parameter. In the limit of unitary scatterers, which is relevant for 
Zn and other solutes in YBCO\cite{Hirschfeld}, the scattering rate is
given by ${1/2\tau}={n_i/\pi N(0)}$,
with $n_i$ being the number density of impurities and 
$N(0)$ the normal-state
density of states at the Fermi level. For $\alpha$ small we have 
\begin{equation}
{T_c^{AG}\over T_{c0}}\simeq
1-{\pi\over 4}\alpha= 1-{n_i\over 4N(0)T_{c0}},
\label{tc-lin}
\end{equation}
which is in fact an excellent approximation to the full solution of (\ref{tc})
for $\alpha\lesssim\alpha_c/3$, where $\alpha_c=0.88191$ is the critical
pair breaking parameter $[T_c(\alpha_c)=0]$.

One of the crucial assumptions entering the derivation of Eq.\ (\ref{tc}) is
that the position dependent order parameter $\Delta({\bf r})$ can be 
replaced in the gap equation by its spatial average $\bar\Delta$. Such a
procedure is valid when the gap varies over a length scale that is large
compared to the average spacing between the impurities, $l_i$.
In the vicinity of $T_c$ significant variations of the order parameter
take place on the length scale set by the Ginzburg-Landau temperature
dependent coherence length $\xi(T)$, which is related to the low temperature
BCS coherence length $\xi_0=v_f/\pi\Delta$ by
$\xi(T)\simeq \nu\xi_0 (1-T/T_c)^{-1/2}$, where $\nu=0.74$ in a conventional
$s$-wave superconductor with a spherical Fermi surface\cite{Tinkham}. For a $d$-wave superconductor $\nu$ will be a different number of
order unity, depending on the precise $k$-space structure of the
gap function.

Consider the 
problem of a single unitary impurity in a $d$-wave superconductor. The order
parameter will be strongly suppressed at the impurity site and it will recover
its bulk value over the distance $\sim\xi(T)$ \cite{Wheatley,Franz}. Since a
single impurity cannot affect $T_c$ of a macroscopic sample, this coherence
length is given by 
\begin{equation}
\xi(T)\simeq \nu\xi_0 (1-T/T_{c0})^{-1/2}.
\label{coh0}
\end{equation}
For a small finite density of impurities with average spacing
$l_i\gg\xi_0$, except very close to $T_c$, the areas of depressed order 
parameter around individual impurities will not overlap, and most of the 
sample will remain completely unaffected. Thus it is still reasonable to use 
Eq.\ (\ref{coh0}) with the unperturbed $T_{c0}$ for $\xi(T)$. When these 
areas begin overlapping, {\em i.e.} when $\xi(T)\gtrsim l_i$,
the entire sample is affected in the sense 
that the order parameter is suppressed everywhere. The temperature at which this happens provides a lower bound for
the true $T_c$ of the sample. Using $l_i=2a_0/\sqrt{\pi n_i}$ 
(where $a_0$ is the ionic lattice spacing) this condition becomes
\begin{equation}
{T_c\over T_{c0}} \gtrsim 1-{\pi\over 4}\left({\nu\xi_0\over a_0}\right)^2
n_i.
\label{cond2}
\end{equation}
If $T_c^{AG}$ predicted from Eq.\ (\ref{tc-lin}) falls below this
lower bound, one may conclude that the AG theory is not self-consistent, since
the true coherence length at $T=T_c^{AG}$ is smaller than the average distance
between the impurities and the averaging over $\Delta({\bf r})$ is not
allowed.

In order to get a rough idea of how restrictive this argument is, we note
that the prefactor of $n_i$ in Eq.\ (\ref{tc-lin}) can be converted into
a length scales ratio, $1/N(0)T_{c0}\sim (\pi^2/1.13)(\xi_0/a_0)$ 
\cite{remark1}. Then, combining 
this with Eq.\ (\ref{cond2})  one can derive a rough estimate for the range of
validity of the AG theory of the form
\begin{equation}
\xi_0/a_0\gtrsim \pi/1.13\nu^2\simeq 5.
\label{range2}
\end{equation}
 Clearly, for conventional low-$T_c$ superconductors this condition is easily 
satisfied as $\xi_0$ is typically large compared to $a_0$.
For high-$T_c$ cuprates however, the situation is quite different since 
typically $\xi_0$ is several lattice spacings 
($\xi_0\approx  4a_0$ in YBCO). 
A more careful estimate  using realistic values of $T_{c0}$ and $N(0)$
for  YBCO
\cite{Arberg}, confirms that the result of our rough estimate (\ref{range2}) 
holds. This indicates
that the usage of the AG theory for this material is probably not 
justified\cite{remark2}.

While the argument presented above is admittedly crude, we believe that
together with the experimental evidence discussed above it points to the
necessity of studying the effects of spatial variations of the order
parameter in disordered short-coherence-length superconductors. We now present
results of a numerical calculation that strongly support the picture
outlined above.

We employ an extended Hubbard model on a square lattice with  
nearest neighbor attraction and on-site repulsion:  
\begin{eqnarray}
H=&-&t\sum_{\langle ij \rangle\sigma} c^\dagger_{i\sigma}c_{j\sigma}   
  -\mu\sum_{i\sigma}n_{i\sigma}
  +\sum_{i\sigma}V^{\rm imp}_i n_{i\sigma}  \nonumber \\
  &+&V_0\sum_i n_{i\uparrow}n_{i\downarrow}
  +{V_1\over 2}\sum_{\langle ij \rangle\sigma\sigma'}n_{i\sigma}n_{j\sigma'},
\label{hub}
\end{eqnarray}
where $\langle ij \rangle$ stands for nearest neighbor pairs, and 
the notation is otherwise standard.  Such a model, treated 
within a self-consistent BdG theory, has been used previously to study vortices \cite{Soininen,Wang} and impurities \cite{Wheatley,Franz,Onischi} 
in $d$-wave superconductors. The impurities are modeled by 
$V_i^{\rm imp}=V^{\rm imp}\gg |t|$ at randomly chosen sites with  density
$n_{i}$ and $V_i^{\rm imp}=0$ elsewhere. We solve this Hamiltonian
within the standard mean-field theory as described in Ref.\ \cite{Franz}. 
All physical quantities of interest  can be derived from the quasiparticle 
amplitudes $[u_n({\bf r}),v_n({\bf r})]$, which satisfy a system of BdG
equations\cite{deGennes}
\begin{equation}
\left( \begin{array}{cc}
\hat{\xi}      & \hat{\Delta} \\
\hat{\Delta}^*  & -\hat{\xi}^*
\end{array} \right)
\left( \begin{array}{c}
u_n \\ v_n
\end{array} \right)
=E_n 
\left( \begin{array}{c}
u_n \\ v_n
\end{array} \right),
\label{BdG}
\end{equation}
with
\begin{eqnarray}
\hat\xi u_n({\bf r}_i) &=& -t\sum_\delta u_n({\bf r}_i+\delta) 
+(V_i^{\rm imp}-\mu)u_n({\bf r}_i), \nonumber \\
\hat\Delta v_n({\bf r}_i)&=&\Delta_0({\bf r}_i) v_n({\bf r}_i)
+\sum_\delta \Delta_\delta({\bf r}_i)v_n({\bf r}_i+\delta),
\end{eqnarray}
 subject to the constraints of self-consistency
\begin{eqnarray}
\Delta_0({\bf r})   &=& V_0\sum_n u_n({\bf r})v_n^*({\bf r}) \tanh(E_n/2k_BT), 
\nonumber \\
\Delta_{\bf \delta}({\bf r}) &=& {V_1\over 2} \sum_n\left[u_n({\bf r}+
{\bf \delta})v_n^*({\bf r})
                                 +u_n({\bf r})v_n^*({\bf r}+{\bf \delta})\right] 
   \nonumber \\
                && \ \ \ \ \ \ \ \ \ \ \ \times \tanh(E_n/2k_BT).
\label{self}
\end{eqnarray}
Here $\Delta_0$ and
$\Delta_\delta$ are the on-site and nearest-neighbor pairing amplitudes
respectively with $\delta=\pm\hat{x}, \pm\hat{y}$ and 
the summation is over positive eigenvalues $E_n$ only. For a finite 
$L\times L$ system we solve the BdG equations (\ref{BdG}) by exact numerical
diagonalization using a suitable guess for the initial gap functions.
We then compute new gap functions from Eq.\ (\ref{self}) and iterate
this process until self-consistency is established\cite{Franz,Soininen}.

The superfluid density $\rho_s(T)$ is evaluated from the standard linear
response formula appropriate for lattice models\cite{Wheatley,Scalapino}
\begin{equation}
\rho_s(T)/4=\langle -K_x\rangle
-\Lambda_{xx}(q_x=0,q_y\to 0,\omega=0),
\label{superfl}
\end{equation}
where $\langle -K_x\rangle$ is the average kinetic energy along the 
$\hat{x}$-direction, and $\Lambda_{xx}({\bf q},\omega)$ is a diagonal
element of the current-current correlation function. Both 
$\langle -K_x\rangle$ and $\Lambda_{xx}({\bf q},\omega)$ can be written in
terms of $(u_n,v_n)$ and $E_n$, resulting in lengthy expressions which
we omit here for brevity.

Determination of $T_c$ is more complicated since the number of iterations
needed to self-consistently solve Eqs.\ (\ref{BdG},\ref{self}) to a required
accuracy increases dramatically near the transition (and presumably diverges
at $T=T_c$). We use two different methods to overcome this difficulty:
{\em (i)} we improve our initial guess for $\Delta({\bf r})$ at each new
temperature by extrapolating from previous temperature points, and
{\em (ii)} we locate $T_c$ iteratively by following the development of 
the gap initialized to infinitesimal value.

\begin{figure}
\epsfxsize=8.0cm
\epsffile{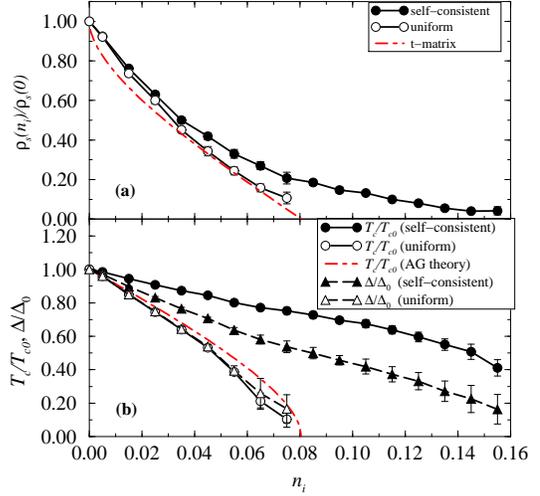}
\caption[]{(a) Normalized zero-temperature superfluid density as a function 
of impurity concentration: fully self-consistent solution of BdG equations
(solid symbols), solution with uniform order parameter (open symbols) and
analytic $t$-matrix solution with unitary scatterers from Ref.\cite{KPM} 
(dash-dotted line).  (b) Normalized critical temperature and average gap. 
Dash-dotted line is numerical solution of the AG equation (\ref{tc}).}
\label{fig:2}
\end{figure}
Figure \ref{fig:2} summarizes the results of our numerical calculation,
for parameters resulting in $\xi_0\approx 4a_0$ (we use $V_0=-V_1=1.13t$,
$\mu=-0.36t$ and $V^{\rm imp}=100t$ for a system size $L=22$) 
Panel (a) shows the normalized zero-temperature superfluid 
density as a function of $n_i$ obtained from a fully self-consistent solution
of Eqs.(\ref{BdG},\ref{self}) and from a solution with an enforced uniform
order parameter. For comparison a conventional analytic solution within
the $t$-matrix formalism is also shown (we use Eq.\ (15) of Ref.\ \cite{KPM}).
As one would expect the uniform solution is very close to the analytical 
one for
all values of $n_i$. The self-consistent solution agrees well with the two,
except for large $n_i$. Our results for $T_c$ [panel(b)] are much more
interesting. While $T_c$ computed for the uniform solution tracks the 
analytic solution of the AG equation (\ref{tc}), the true critical temperature
obtained using a self-consistent solution is much higher. In fact at 
$n_i\simeq 0.08$ where $T_c^{AG}=0$, the true critical temperature is still
more than 70\% of $T_{c0}$. Also note that the average zero-temperature
gap does not scale with $T_c$, as one would expect from BCS theory. 

\begin{figure}
\epsfxsize=8.0cm
\epsffile{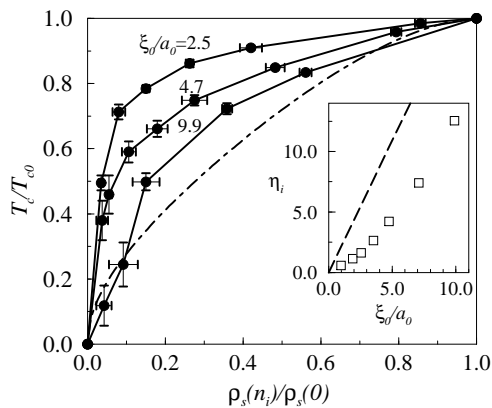}
\caption[]{Normalized critical temperature versus
normalized zero-temperature superfluid density as computed from a 
fully self-consistent solution of BdG equations for systems with 
different coherence lengths. The error bars reflect the statistical scatter
of data for 6 different impurity configurations. Parameters used are
$L=22$, $V^{\rm imp}=-100t$, $\mu=-0.36t$, and $V_0=-V_1=(0.80,1.05,1.40)t$ for
$\xi_0=(9.9,4.7,2.5)a_0$ respectively. Dash-dotted line is a conventional 
AG solution from Ref.\ \cite{KPM}. Inset: the rate of change $\eta_i$
versus $\xi_0/a_0$ (open squares), and the expected behavior from AG theory
$\eta_i=2.18(\xi_0/a_0)$ (dashed line).}
\label{fig:3}
\end{figure}

We have carried out 
calculations for several different sets of parameters, all showing similar
behavior (see Fig.\ \ref{fig:3}). As the coherence length grows, the 
discrepancy between the true 
critical temperature and $T_c^{AG}$ diminishes, as expected from the 
analysis presented above. For $\xi_0/a_0\approx 4$ as in 
YBCO, the discrepancy is 
substantial, and can easily account for the experimentally observed 
robustness of $T_c$. The critical impurity density for this case is 
$n_i^c\approx 0.17$. Since Zn substitutes primarily for the planar 
copper sites in YBCO\cite{Villeneuve} and only 2/3 of all Cu atoms reside 
in the planes, our result implies a bulk critical density of 
$n^c_{\rm bulk}\approx 0.10$. 
This is in reasonable agreement with the experimentally observed 
$n^c_{\rm bulk}=0.08-0.10$ for Zn doped YBCO\cite{Ishida}.
By contrast conventional models tend to 
underestimate $n^c_{\rm bulk}$ by a factor of 2 or more\cite{Fehrenbacher}.
The inset to Fig.\ \ref{fig:3} shows the dependence of the initial rate of
change of $T_c$ with $n_i$, $\eta_i=T_{c0}^{-1}(dT_c/dn_i)_{n_i=0}$, on
$\xi_0$. This quantity also deviates significantly from the linear
$\eta_i\sim (\xi_0/a_0)$ behavior expected from the AG prediction
(\ref{tc-lin}), showing instead a quadratic behavior consistent with 
(\ref{cond2}).

In closing we comment on the appropriateness of using BdG theory to
calculate $T_c$ in short coherence length materials.  Like all conventional
theories of superconductivity, BdG theory is a mean field theory and hence
is incapable of modeling the important effects of phase fluctuations on
the ordering temperature in short coherence length, highly anisotropic
materials such as the high-$T_c$ oxides.  On the other hand, unlike many
more specialized mean field theories, including AG theory, BdG theory is
well-suited to calculating the effects of static spatial disorder on the
order parameter amplitude and on the mean field $T_c$.  Since the static
random potential does not couple to the phase, the effects of spatial
randomness and thermal phase fluctuations are largely decoupled.  To a
good approximation, at least for the dilute case, the effects of
impurities can be calculated within mean field theory, while the effects
of phase fluctuations, in reducing $T_c$ from its mean field value,
require a proper theory of critical behavior.  A complete theory for the
combined effects is not presently known.  However it is apparent, even at
the mean field level, that such a theory for the effect of inhomogeneous
pair-breaking on $T_c$ must take proper account of the resulting
inhomogeneity of the gap function.  It is our hope that the present
calculations will provide the physical motivation for such a theory.

The authors are indebted to D. N. Basov, C. Bernhard, H. Kim and S. H. Moffat
for providing their data for Fig.\ \ref{fig:1}, and are grateful to 
A. V. Balatsky, V. J. Emery, D. L. Feder, R. Fehrenbacher, K. Levin, 
J. S. Preston, and T. Timusk for inspiring discussions.
This work has been partially supported by NSERC, OCMR 
and by NSF Grant No. DMR-9415549 (M.F.).


\begin{references}
%
\bibitem[*]{} Permanent address: Department of Physics and Astronomy,
McMaster University,  Hamilton, Ontario, L8S 4M1 Canada. 
\bibitem{Annett} J. F. Annett, N. D. Goldenfeld and S. R. Renn, 
\prb{\bf 43}, 2778 (1991).
\bibitem{Hirschfeld} P. J. Hirschfeld and N. D. Goldenfeld, 
\prb{\bf 48}, 4219 (1993).
\bibitem{Hardy} W. N. Hardy {\em et al.}, \prl{\bf 70}, 3939 (1993).
\bibitem{Prohammer} M. Prohammer and J. P. Carbotte, \prb{\bf 43}, 5370 (1991).
\bibitem{KPM} H. Kim, G. Preosti and P. Muzikar, \prb{\bf 49}, 3544 (1994).
\bibitem{Maki} Y. Sun and K. Maki, \prb{\bf 50}, 6059 (1995).
\bibitem{Radtke} R. J. Radtke {\em et al.} \prb{\bf 48}, 653 (1993).
\bibitem{AG} A. A. Abrikosov and L. P. Gorkov, Zh.\ Eksp.\ Teor.\ Fiz.\ 
{\bf 39}, 1781 (1960) [Sov.\ Phys.\ JETP {\bf 12}, 1243 (1961)].
\bibitem{Ulm} E. R. Ulm {\em et al.} \prb{\bf 51}, 9193 (1995).
\bibitem{Basov} D. N. Basov {\em et al.}, (unpublished).
\bibitem{Basov1} D. N. Basov {\em et al.}, \prb{\bf 49} 12165 (1994).
\bibitem{Tallon} C. Bernhard {\em et al.}, \prl{\bf 77}, 2304 (1996).
\bibitem{Moffat} S. H. Moffat {\em et al.}, (unpublished).
\bibitem{Arberg} P. Arberg and J. P. Carbotte, \prb{\bf 50}, 3250 (1994).
\bibitem{Fehrenbacher} R. Fehrenbacher, \prl{\bf 77}, 1849 (1996).
\bibitem{Pines} P. Monthoux and D. Pines, \prb{\bf 49}, 4261 (1994).
\bibitem{Hirschfeld2} P. J. Hirschfeld, J. Phys.\ Chem.\ Solids {\bf 56}, 
1605 (1995).
\bibitem{Tinkham} M. Tinkham, {\it Introduction to Superconductivity} 
(Krieger, Malabar, 1975).
\bibitem{Wheatley} T. Xiang and J. M. Wheatley, \prb{\bf 51}, 11721 (1995).
\bibitem{Franz} M. Franz, C. Kallin, and A. J. Berlinsky, \prb{\bf 54},
R6897 (1996).
\bibitem{remark1} Here we use $1/N(0)\approx\pi v_f/2a_0$, 
valid for a tight-binding model away from half filling and from the bottom of 
the band, and a BCS relation $T_{c0}/\Delta\simeq 1.13/2$. For a $d$-wave 
superconductor this ratio will depend on the structure of the gap function, 
but will remain of order one. For concreteness and simplicity we use the 
BCS result. 
\bibitem{remark2} In a different context the breakdown of the AG theory has 
been discussed by Balatsky, Salkola and Rosengren 
[\prb {\bf 51}, 15547 (1995)].
\bibitem{Soininen} P. I. Soininen, C. Kallin and A. J. Berlinsky, 
\prb{\bf 50}, 13883 (1994).
\bibitem{Wang}Y. Wang and A. H. MacDonald, \prb{\bf 52}, R3876 (1995).
\bibitem{Onischi} Y. Onischi, {\em et al.} J. Phys.\ Soc. Jpn.\ {\bf 65}, 
675 (1996).
\bibitem{deGennes}P. G. de Gennes, {\em Superconductivity of Metals and 
Alloys}, (Addison-Wesley, Readings, MA, 1989).
\bibitem{Scalapino} D. J. Scalapino, S. R. White and S. C. Zhang, 
\prl{\bf 68}, 2830 (1992).
\bibitem{Villeneuve} R. Villeneuve {\em et al.}, Physica C {\bf 235-240}, 
1567 (1994).
\bibitem{Ishida} K. Ishida {\em et al.}, J. Phys.\ Soc.\ Jpn.\ {\bf 62}, 
2803 (1993).

\end{references}
\end{document}